\documentclass[12pt]{iopart}
\usepackage{graphicx}
\begin{document}
\title{Molecules with dipoles in periodic boundary
  conditions in a tetragonal cell}

\author{M J Rutter}

\address{TCM, Cavendish Laboratory, JJ Thomson Avenue, Cambridge,
  CB3~0HE, UK}

\begin{abstract}
  When a system which contains a dipole, and whose dimensionality is
  less than three, is studied in a code which imposes periodic boundary
  conditions in all three dimensions, an artificial electric field
  arises which keeps the potential periodic. This has an impact on
  the total energy of the system, and on any other attribute which
  would respond to an electric field. Simple corrections are known for
  0D systems embedded in a cubic geometry, and 2D slab systems. This
  paper shows how the 0D result can be extended to tetragonal
  geometries, and that for a particular $c/a$ ratio the correction is zero. It
  also considers an exponential error term absent from the usual
  consideration of 2D slab geometries, and discusses an empirical form
  for this.
\end{abstract}

\maketitle

\section{Introduction}

Plane-wave electronic structure codes enforce periodicity in all three
dimensions. Whilst this is ideal for studying bulk crystals, it can
cause complications when the system of interest has lower
dimensionality, such as a surface (2D), a nanowire (1D), or an
isolated molecule (0D). In order to study these systems in such codes,
a region of vacuum is used to separate the system from its fictitious
periodic images. As the extent of the vacuum is increased, properties
converge to the values that they would have in the absence of the
imposed periodicity.

Large amounts of vacuum significantly increase the memory and time
costs of calculations, so it is useful to be able to apply corrections
which accelerate the convergence with vacuum size. This paper
considers the interactions arising from dipole--dipole interactions in
charge-neutral systems. In such systems, this interaction decays the
most slowly with system size.

For the 2D geometry of a slab of material, commonly used to study
surfaces, a dipole moment perpendicular to the slab results in a
compensating electric field in order to keep the potential
periodic\cite{Scheffler92}. A 2D slab with a perpendicular dipole
moment is effectively a charged parallel plate capacitor, and one
would expect the potential on each side to differ, and the field
outside the plates to be zero. This is not consistent with the
potential being zero at infinity, so a compensating field arises
automatically.

A simple, effective, but computationally-expensive, solution to this
problem is to use a simulation cell consisting of two slabs in
opposite orientations so that the total dipole moment is zero.

A self-consistent correction adds an equal and opposite field within
the simulation. This can be done by placing the required discontinuity
in the potential in the vacuum region\cite{Scheffler92,Bengtsson99}.
Many DFT codes include this approach\cite{CASTEP,QE17,VASP}.

A \textit{post hoc} correction simply involves adding a term to the
final energy. This make no improvement to the convergence of other
properties of the system, such as dipole moments or bond lengths, but
does significantly improve the energy. The energy correction is

\begin{equation} \label{eq:Eslab}
  \Delta E_{\textrm{slab}} = \frac{p^2}{2\epsilon_0V} 
\end{equation}

where $p$ is the dipole moment, which must be perpendicular to the
slab, and $V$ the unit cell volume\cite{Bengtsson99}. The electric
field arising is simply $\mathbf{p}/\epsilon_0V$. These corrections
decay as the reciprocal of the slab separation. These three schemes
are discussed in more detail in reference~\cite{Natan06}.

Similar expressions arise for the 0D geometry of a molecule in a cubic
box\cite{NdW58,MP95}, save that they have an extra prefactor of a
third,

\begin{equation} \label{eq:Ecube}
  \Delta E_{\textrm{cube}} = \frac{p^2}{6\epsilon_0V} 
\end{equation}

and a field correction of $\mathbf{p}/(3\epsilon_0V)$. These
corrections for the cubic geometry are referred to later as the
Makov-Payne correction\cite{MP95}.

More general geometries have been considered\cite{NdW58,MP95,Kant99},
and extended to systems embedded in anisotropic dielectrics\cite{Turban16},
but these tend to result in a requirement to perform a infinite lattice sum,
rather than a simple rational prefactor.

Other methods have been considered to correct for unwanted
dipole-dipole interactions. These include the truncation of the
Coulomb potential in real space\cite{Jarvis97}, which was proven to be
equivalent to the above self-consistent slab
correction\cite{Yu08}. However, Coulomb truncation methods require the
extent of the vacuum region to be greater than the extent of the
non-vacuum region\cite{Rozzi06}. Sharp cut-offs, and `minimum image'
modifications, to the Coulomb potential are compared in detail in
Ref~\cite{Hine11}.

For the slab geometry a different approach is to use a 2D version of
the Ewald sum\cite{Parry75}. This has also shown to be equivalent to
the 3D Ewald sum with the above slab correction\cite{Brodka02}.

This paper considers a limited generalisation of the cube result of
Makov and Payne, a generalisation which avoids the need for
summations, and it considers a particular geometry for which no
correction is needed.  It also finds a further correction term for the
2D slab geometry.

\section{Tetragonal Cells}

The corrections described above relate calculation cells to an ideal
cell in which one, or more, of the repeat distances has been extended
to infinity. The geometry for which further results can be readily
obtained is the 0D system of a molecule in a tetragonal cell with the
molecule's dipole moment parallel to the $\mathbf{c}$ axis, the cell volume
being $V=a^2c$.

Equation~\ref{eq:Eslab} allows one to calculate the energy change
arising from the dipole interaction between the periodic images of the
simulation cell if $c$ were increased to infinity. It also allows one
to calculate what the energy would be in another simulation cell with
a different value of $c$, by first increasing to infinity, and then
reducing to the new value. So one can consider what the dipole-dipole
interaction energy would have been had the calculation been performed
in a cubic cell, one with $c=a$, albeit ignoring the warning given
by Yeh and Berkowitz\cite{Yeh99} that the slab correction loses accuracy unless
the vacuum gap size exceeds $a$. The correction would be:

\begin{equation}
  \Delta E = \frac{p^2}{2\epsilon_0a^2c} -  \frac{p^2}{2\epsilon_0a^3}
\end{equation}

Now that the cell is cubic, one can apply equation~\ref{eq:Ecube} to
expand $a$ to infinity in a cubic geometry, giving

\begin{equation}
  \Delta E = \frac{p^2}{2\epsilon_0a^2c} - \frac{p^2}{2\epsilon_0a^3} + \frac{p^2}{6\epsilon_0a^3}
\end{equation}

which simplifies to

\begin{equation}
  \Delta E_{\textrm{tetragonal}} =
  \frac{p^2}{\epsilon_0a^2}\left(\frac{1}{2c} - \frac{1}{3a}\right)
  \label{eq:tet}
\end{equation}

This is thus the energy correction arising from dipole-dipole
interactions required to expand a charge-neutral, finite, tetragonal
cell with the dipole moment wholly along the $\mathbf{c}$ axis to
infinite size.  It is zero if
$c=1.5a$.

That the energy correction is zero implies that the correcting field
that should be added in a self-consistent calculation will also be
zero, and repeating the above calculations for the field rather than
the energy confirms this.

So by careful choice of the 3D simulation cell geometry one can cause
the dipole-dipole interaction for a 0D system to cancel, even in a
code with no built-in correction.

\section{Calculations}

In order to demonstrate the above theory, some calculations were
performed on a KCl molecule using \textsc{Castep}\cite{CASTEP}, and
the dipole moments and \textit{post hoc} corrections calculated with
c2x\cite{c2x}. As only the convergence of properties with cell size
was of interest, little thought was given to the
choice of exchange-correlation functional or pseudopotential, save
that a single-electron pseudopotential was used for K for its speed
benefits.  The bond length was fixed at 2.7\AA.

\begin{figure}
  \begin{center}
    \includegraphics[width=0.6\textwidth]{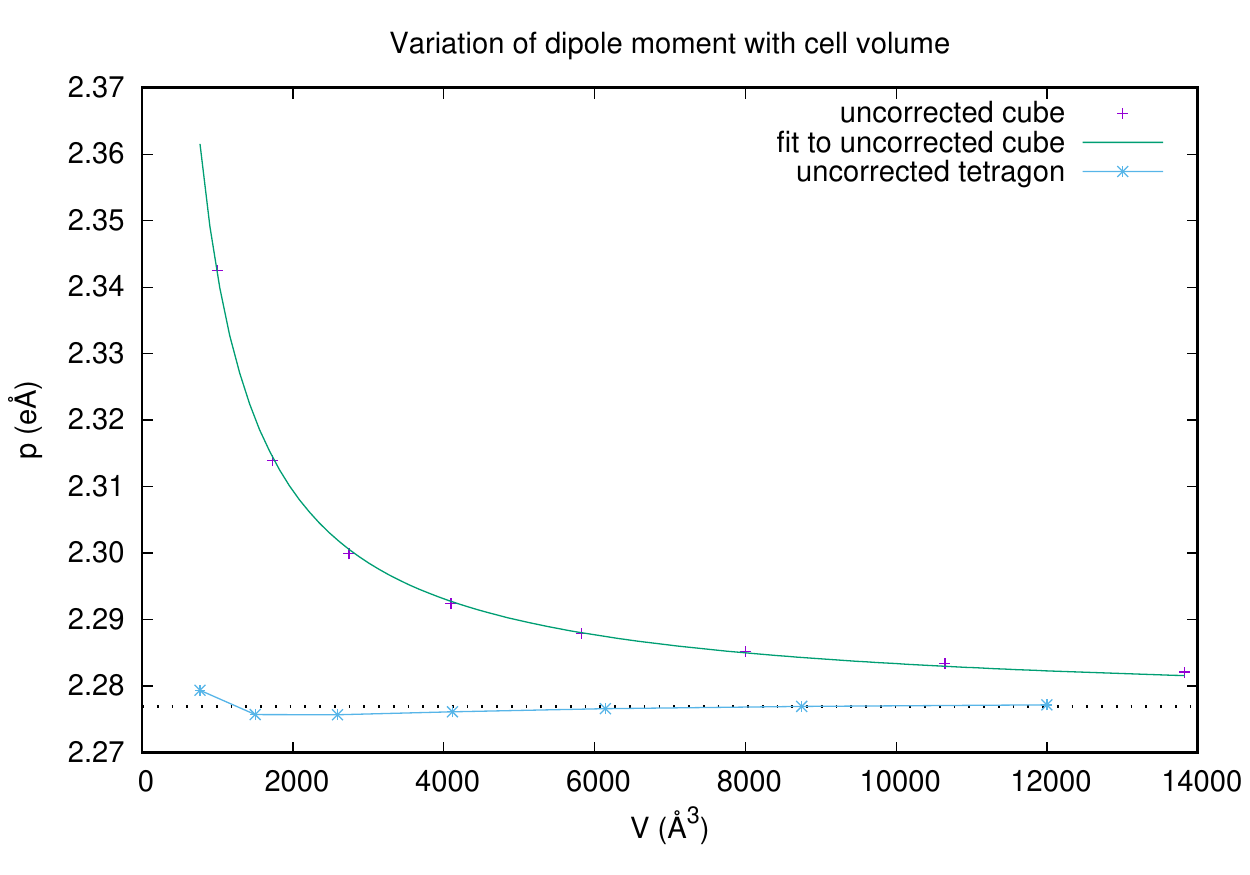}
  \end{center}
  \caption{Variation of dipole moment of KCl molecule in a cubic
    box, and a tetragonal box with $c=1.5a$. For the cubic box, the
    crosses represent calculations, and the line is a
    best-fit of constant plus a decay inversely proportional to
    volume. The horizontal dotted line is the asymptote of this fit.}
  \label{fig:dipole}
\end{figure}

First the dipole moment of the KCl molecule was considered. In a cubic
cell the effect of the periodic images will be to produce a field
which enhances the dipole moment through polarisation, and this field
should decay as the reciprocal of the cell volume. In a tetragonal
cell with $c=1.5a$, this field is expected to be zero, and therefore
the convergence with cell size is expected to be much faster.

The results from uncorrected cubic cells are shown in
figure~\ref{fig:dipole}, together with a fitted line of the form
$\alpha+\beta/V$ where $V$ is the cell volume. The value of $\alpha$
is 2.27688e\AA{}. The same calculation was performed in tetragonal
boxes with $c=1.5a$ and $c$ being 15, 18, 24 and 30\AA. The results from
the tetragonal cell show very little variation with cell volume, and lie
close to the value obtained by extrapolating the curve fitted to the cubic
data to infinite volume. A 10\AA{}x10\AA{}x15\AA{} tetragonal box
gives about one fifth of the error of the cube of side 24\AA{}, in a
cell of little more than one tenth of the volume.

\begin{figure}
  \begin{center}
    \includegraphics[width=0.6\textwidth]{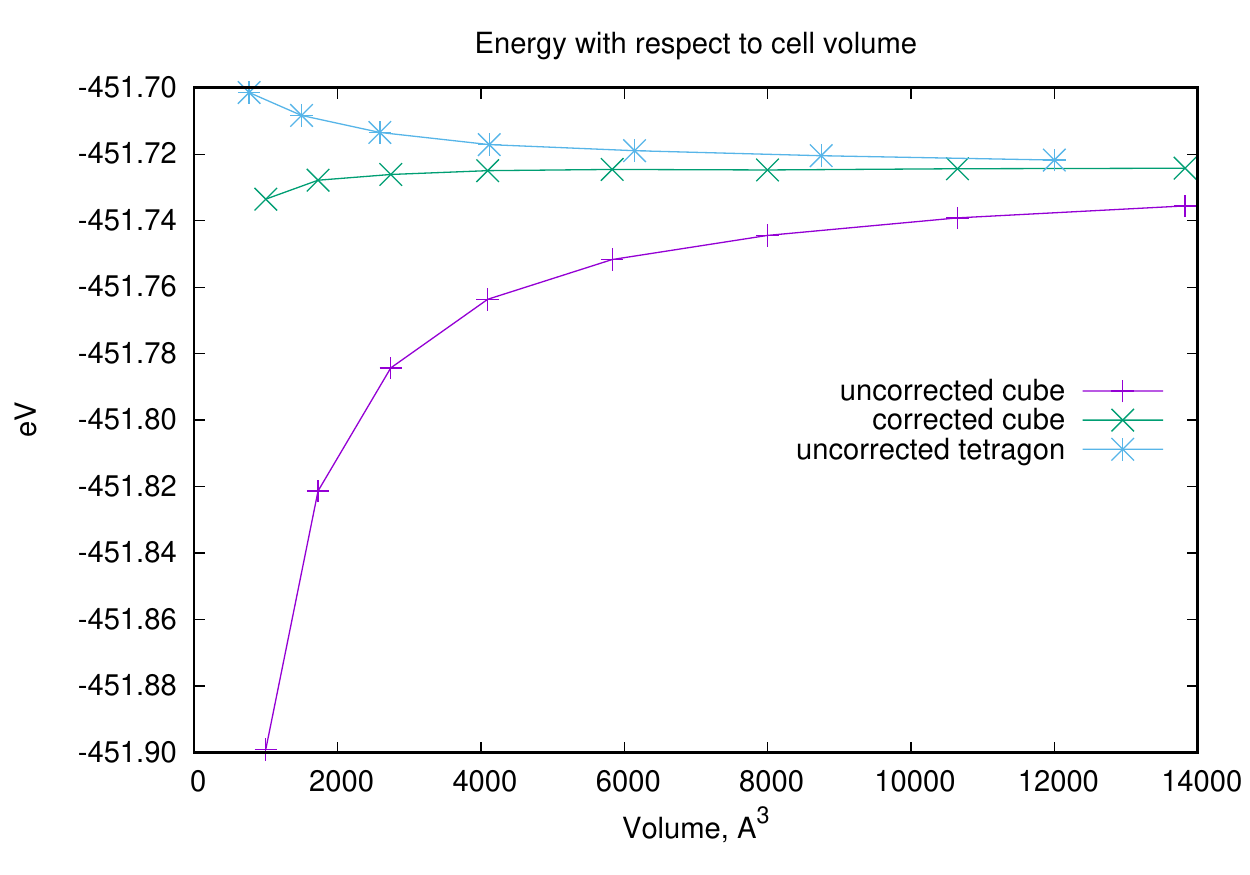}
  \end{center}
  \caption{Variation of the energy of a KCl molecule in cubic
    and tetragonal (c=1.5a) boxes of various volumes.}
  \label{fig:energy}
\end{figure}

The energy convergence is also improved by the use of a tetragonal
cell, but is found to be less good than a cubic cell with
the correction of equation~\ref{eq:Ecube}. This is shown in
figure~\ref{fig:energy}, where the x-axis is now volume to give a fairer
comparison between cubic and tetragonal cells. Whilst the use of a tetragonal
cell is a significant improvement over an uncorrected cubic cell,
attempts to fit equations to the remaining error show that a small
term inversely proportional to the volume remains.

\begin{figure}
  \begin{center}
    \includegraphics[width=0.6\textwidth]{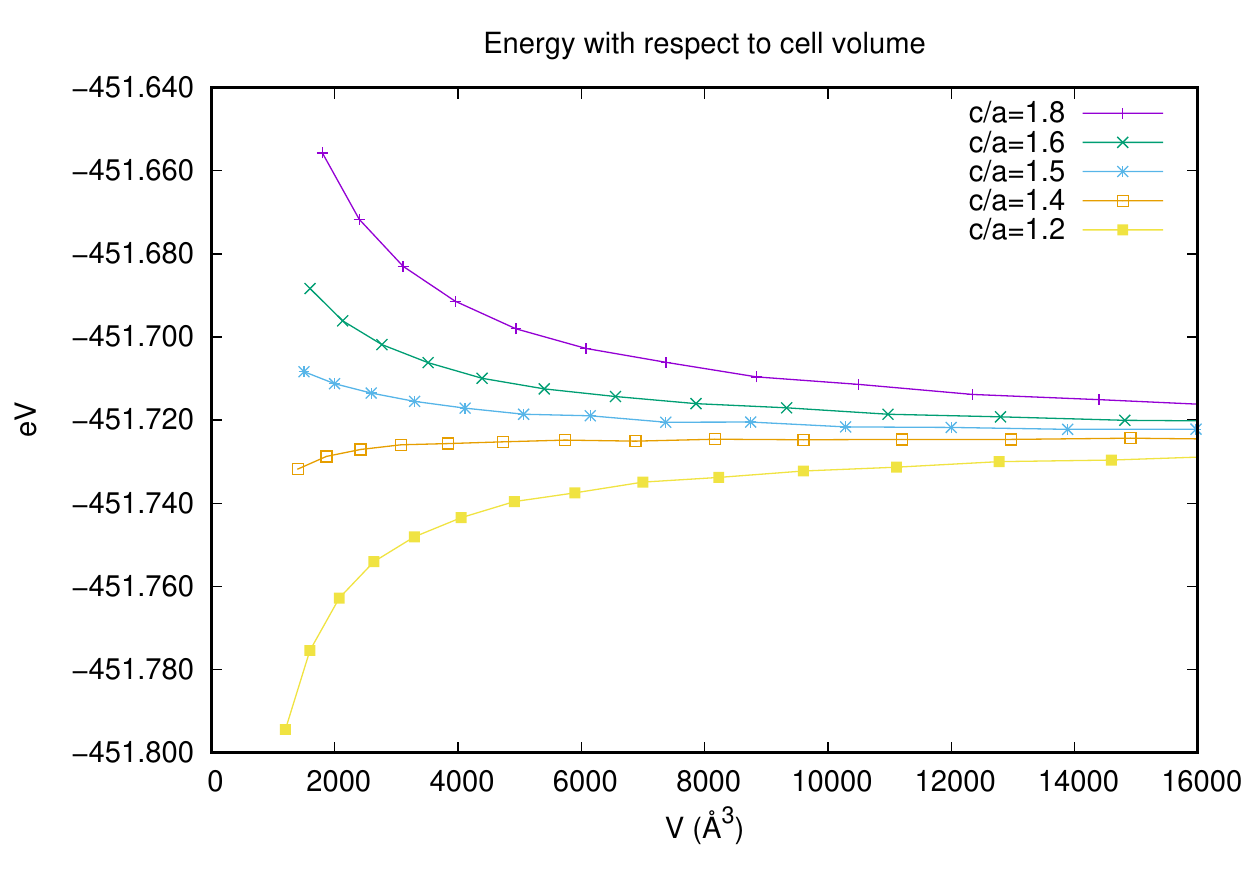}
  \end{center}
  \caption{Variation of the energy of a KCl molecule in
    tetragonal boxes of various $c/a$ ratios and different volumes.}
  \label{fig:energy-tet}
\end{figure}

If the variation of energy at different $c/a$ ratios is considered, as
presented in figure~\ref{fig:energy-tet}, then the movement from the
dominant term being a lowering of energy from the attractive
interactions in the $c$ direction at low values of $c/a$ to the rise
in energy as the repulsive interactions perpendicular to $c$ at higher
values of $c/a$ is clear. However, cancellation of these effects
appears to occur at a value of $c/a$ only very slightly greater than 1.4,
and not at 1.5 as this section would suggest. The following section
explores the reason for this.

\section{The Slab Correction Revisited}

The theory of the slab correction is usually expressed in terms of
parallel plate capacitors. In the periodic system, the planes half-way
between the slabs have constant potential. If a thin conducting sheet
were placed in one such plane, then the two halves of the system would
no longer be influenced by each other, but each would see in the sheet
a set of mirror charges which would be identical to the now-shielded
half of the system. Hence such conducting sheets may be introduced
without changing the system.

A capacitor of plate separation $d$ and surface charge density
$\sigma$ has a dipole moment per unit area of $\sigma d$, a field
between the plates of $\sigma / \epsilon_0$, and a potential
difference of $\sigma d /\epsilon_0$. So it is argued that in the
simulation cell a potential difference of $p / \epsilon_0 A$ would be
expected, where $p$ is the dipole moment perpendicular to the slab,
and $A$ the cell's area in the plane of the slab. That periodic
boundary conditions apply to the potential forces the removal of this
potential difference by the imposition of an artificial field of
magnitude $p / \epsilon_0 V$. The corrections of
equation~\ref{eq:Eslab} and the corresponding field correction account
for this unwanted electric field.

This leaves many terms unaccounted for: dipole moments parallel to the
slab, quadrapole and higher moments, polarisation effects, and effects
from the finite extent of dipoles as the theory applies to point
dipoles. However, there are also terms arising purely from the static
dipole which are not properly addressed.

\begin{figure}
  \begin{center}
    \includegraphics[width=0.6\textwidth]{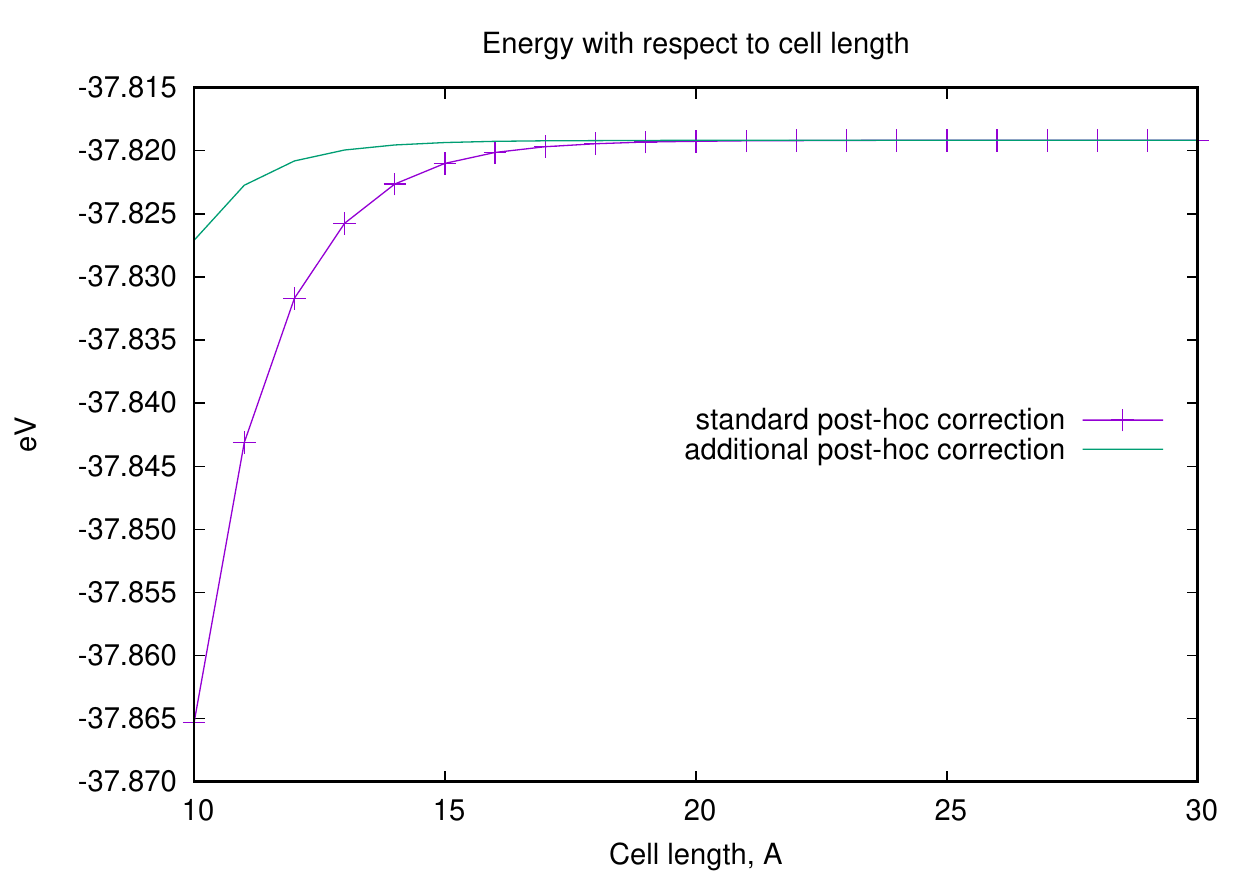}
  \end{center}
  \caption{Variation of the post-hoc corrected Ewald slab energy with
    varying slab separation for a system of two point charges
    separated by 1.5\AA{} in the
    centre of a teragonal cell with a=10\AA{} and c varying. The
    line with points shows calculated values, and the bare line shows
    the result of adding the second correction proposed in
    equation~\ref{eq:Ecorr}.}
  \label{fig:Ewald}
\end{figure}

It is instructive to consider a simpler system than the KCl molecule
of the previous system, that of the Ewald energy of two equal and
opposite point charges in a cell. The Ewald routine from the python
package pymatgen\cite{pymatgen} was used for this. The system was two
charges of $+2e$ and $-2e$ separated by 1.5\AA{} in a tetragonal box
with $a=$10\AA{} and $c$ varying from 5\AA{} to 50\AA. The energy of
this system for part of this range, corrected by
equation~\ref{eq:Eslab}, is shown in figure~\ref{fig:Ewald}. Also shown
in the figure is the result of adding a second empirical correction term of

\begin{equation}
\delta E = \frac{4 \pi p^2}{\epsilon_0a^3}\exp(-2\pi c/a) \label{eq:Ecorr}
\end{equation}

This correction was found to fit data well for various configurations of
the test system, with the fit improving as the separation of the
charges was reduced. Whilst the author is unable to offer analytic
proof of the above form, the following observations can be made.

The electrostatic potential in a plane above and parallel to an
isolated square array of dipoles is not a constant. Close to the slab
it will vary as one moves from being close to a dipole in the slab to
being between dipoles. The arguments based on capacitors assume that
the potential is constant, as it would be if the dipole moment were
uniformly distributed across the area of the slab, and also as it is
far from the slab. Any variation in the potential $\phi$ in the vacuum
region must obey Laplace's equation of $\nabla^2\phi=0$. The variation
parallel to the slab will have an oscillatory form with the longest
wavelength component being $\exp(2\pi i x/a)$, leading to a decay away
from the slab of the form $\exp(-2 \pi z/a)$. This is the
justification offered for the $\exp(-2\pi c/a)$ term in
equation~\ref{eq:Ecorr}.

Dimensionality suggests the factor of $p^2/\epsilon_0$ and also the
reciprocal of a volume. That the volume appears as $1/a^3$ not
$1/a^2c$ is merely because that fitted the data very much better. The
final factor of $4\pi$ arises simply from coercing a fitted constant
to the nearest likely number.

Other exponentially decaying terms are to be expected,
corresponding to shorter-wavelength components of the variation of the
potential parallel to the plane, but the term quoted above will be
the most slowly decaying. If the dipole were smeared out in the plane of
the slab, and thus had the geometry of a parallel plate
capacitor, then this correction term would not apply.

The same Ewald calculations with pymatgen can be performed in cubic
cells. In this case the residual error after applying the traditional
Makov-Payne correction was well fitted by a function decaying as
$1/a^5$ as expected.

\subsection{Returning to the Tetragonal Cell}

When the slab geometry is considered, the additional correction
proposed in equation~\ref{eq:Ecorr} decays exponentially with cell
length. In the case of a tetragonal cell in which $c/a$ is fixed as
the cell size is varied, it merely decays as $1/V$. For the two shapes
of tetragonal cell considered in this paper, that of $c=a$ (a cube),
and $c=1.5a$, it takes the approximate values

\begin{figure}
  \begin{center}
    \includegraphics[width=0.6\textwidth]{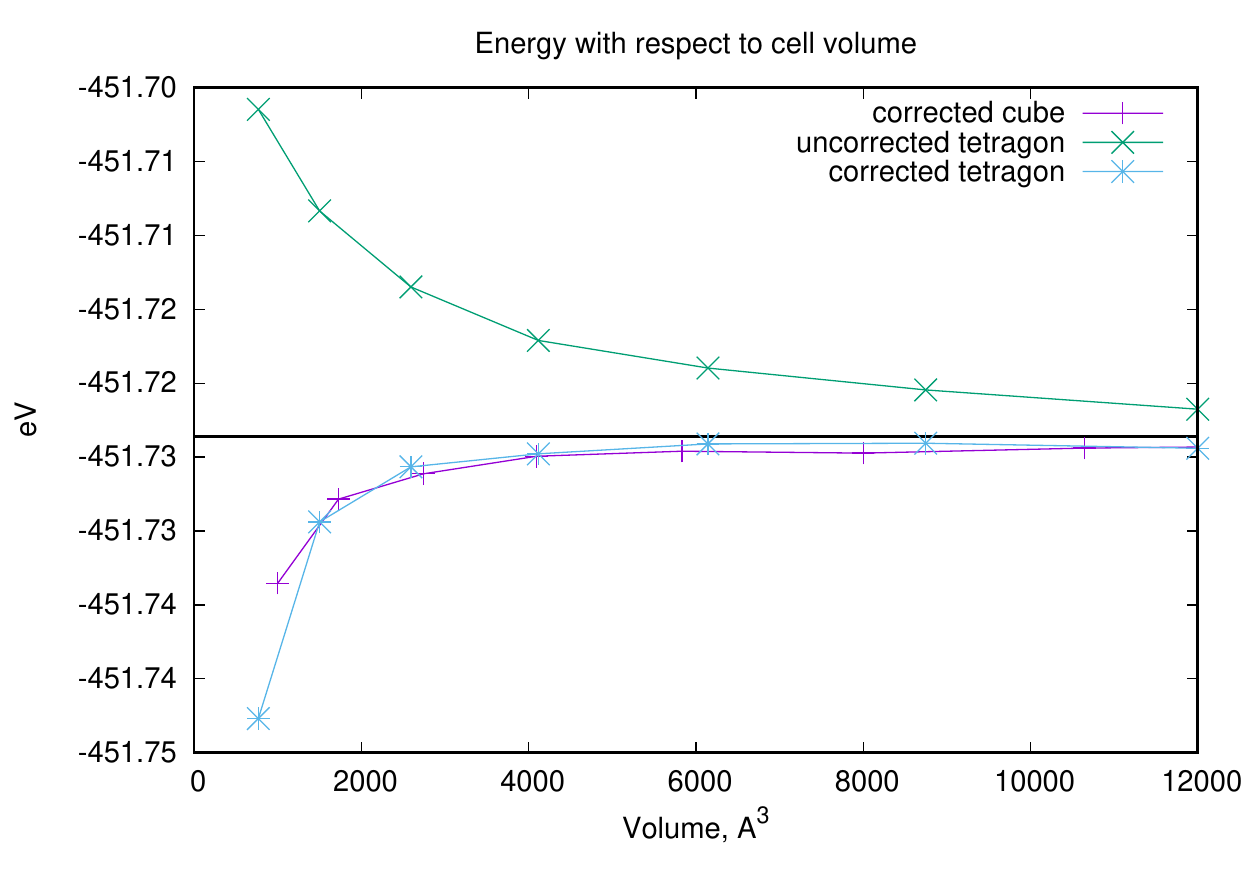}
  \end{center}
  \caption{Variation of the energy of a KCl molecule in 
    tetragonal (c=1.5a) boxes of various volumes, with and without the
    correction term proposed in this paper. The horizontal line is a
    guide to an extrapolation of the two datasets. The corrected cubic
  result from figure~\ref{fig:energy} is given for comparison, and
  lies above the corrected tetragonal result at 1000\AA{}$^3$}
  \label{fig:energy2}
\end{figure}

\begin{equation} \label{eq:dEcube}
  \delta E_{\textrm{cube}} = 0.0235 \frac{p^2}{\epsilon_0V} 
\end{equation}

and

\begin{equation} \label{eq:dEtetra}
  \delta E_{\textrm{tetragonal}} = 0.0015 \frac{p^2}{\epsilon_0V} 
\end{equation}

The arguments from equation~\ref{eq:tet} which stated that there was
no energy correction term for $c/a=1.5$ neglected this correction, and
so there does remain a correction of

\begin{equation}
\delta E = \frac{4 \pi p^2}{\epsilon_0a^3}\left(\exp(-2\pi)-\exp(-3\pi)\right)
\label{eq:tet2}
\end{equation}

This can now be applied to figure~\ref{fig:energy} to produce
figure~\ref{fig:energy2}. This shows that such a term improves
convergence, and gives very similar results to those obtained from a
cubic cell with its \textit{post hoc} correction. Given that
equation~\ref{eq:tet} relies on equation~\ref{eq:Ecube} for the
\textit{post hoc} correction of a cubic cell, it would not
be expected to perform better than this.

The leading term in the error in equation~\ref{eq:Ecube} arises from
the non-zero length of the dipole which produces a term which decays
as $1/a^5$. The energy of an isolated dipole of two charges $\pm q$
separated by $r$ is $-q^2/(4\pi\epsilon_0r)$, or
$-p^2/(4\pi\epsilon_0r^3)$. This term becomes large as the separation
is reduced at constant dipole magnitude and dominates the Ewald
sum. The finite precision of double precision arithmetic then prevents
precise calculation of the interaction energies between adjacent
dipoles by considering the differences in the Ewald energies of the systems.

Figure~\ref{fig:energy-tet} showed that the best convergence of energy
with respect to $c/a$ ratio was found at $c/a$ just over 1.4, and not
at exactly 1.5. However, the best $c/a$ ratio for converging the
energy is not expected to be found by setting equation~\ref{eq:tet} to
zero and solving

\begin{equation}
  \frac{p^2}{\epsilon_0a^3}\left(\frac{a}{2c} - \frac{1}{3}\right)=0
\end{equation}

but rather by considering also the correction of
equation~\ref{eq:Ecorr} and solving

\begin{equation}
  \frac{p^2}{\epsilon_0a^3}\left(\frac{a}{2c} - \frac{1}{3}\right)
  -\frac{4 \pi p^2}{\epsilon_0a^3}\left(\exp(-2\pi)-\exp(-2\pi c/a)\right)=0
\end{equation}

This has a numerical solution of $c/a \approx 1.4085$, which is
consistent with figure~\ref{fig:energy-tet}. However, this $c/a$ ratio
will not result in no correcting electric field, and thus will not
produce the best results for the dipole moment, nor for other properties
which will be influenced by such a field.

\section{Conclusions}

The well-known analytic corrections for dipole-dipole interactions for
a molecule in a cubic box with periodic boundary conditions have been
generalised slightly to include tetragonal boxes, as long as the
dipole moment is parallel to the $\mathbf{c}$ axis. This
generalisation shows that for an $c$ to $a$ ratio of $3/2$ the standard
corrections for both energy and internal fields vanish. This provides
a way of performing accurate calculations when implementing the full
self-consistent correction may be impractical. It may be found to be
of particular use for studying dipolar defects in a dipole-free
medium.

An additional correction term is proposed for systems consisting of a
point dipole in a slab geometry. However, it should be noted that this
correction does not apply if the dipole is smeared out in the plane of
the slab, as is often the case when studying surfaces. When it does
apply, the $c/a$ ratio required to produce the best energy convergence
of a point dipole in a tetragonal cell is reduced from $3/2$ to
approximately 1.41.

\section{Acknowledgements}

This work was supported by EPSRC grant number EP/P034616/1.

\section*{References}

\bibliographystyle{iopart-num}
\bibliography{dipoles}

\end{document}